\documentclass[a4paper, amsfonts, amssymb, amsmath, reprint, showkeys, nofootinbib, twoside]{revtex4}
\usepackage{graphicx}
\usepackage{bm,natbib}
\usepackage{amsmath}
\usepackage{amssymb}
\usepackage{amsfonts}
\usepackage{epsfig}
\usepackage{colordvi}
\usepackage{psfrag}
\usepackage{color}
\usepackage{ mathrsfs }
\newcommand{\Lagr}{\mathcal{L}}

\newcommand{\R}{\mathcal{R}}
\newcommand{\G}{\mathcal{G}}

\newcommand{\La}{\mathscr{L}}
\usepackage{dcolumn}
\usepackage{multirow}
\usepackage{hyperref}
\usepackage{epstopdf}
\usepackage{bm}
\usepackage{rotating}
\usepackage{lscape}
\usepackage{times}
\usepackage{comment}

\hypersetup{
  colorlinks=true,        % false: boxed links; true: colored links
  linkcolor=blue,         % color of internal links
  citecolor=magenta,      % color of links to bibliography
}
\usepackage[english]{babel}
\usepackage[utf8]{inputenc}
\usepackage[colorinlistoftodos, color=green!40, prependcaption]{todonotes}

\begin{document}

\title{The Noether Symmetry Approach: Foundation and applications. \\ The case of scalar-tensor  Gauss-Bonnet gravity}
\author{Francesco Bajardi}
   \email{f.bajardi@ssmeridionale.it}
\affiliation{Scuola Superiore Meridionale, Largo San Marcellino 10, I-80138, Napoli, Italy.}
\affiliation{INFN Sez. di Napoli, Compl. Univ. di Monte S. Angelo, Edificio G, Via Cinthia, I-80126, Napoli, Italy.}
\author{Salvatore Capozziello}
\email{capozziello@na.infn.it}
\affiliation{Dipartimento di Fisica ``E. Pancini'', Universit\'a di Napoli ``Federico II'',  Compl. Univ. di Monte S. Angelo, Edificio G, Via Cinthia, I-80126, Napoli, Italy.}
\affiliation{Scuola Superiore Meridionale, Largo San Marcellino 10, I-80138, Naples, Italy.}
\affiliation{Istituto Nazionale di Fisica Nucleare (INFN) Sez. di Napoli, Compl. Univ. di Monte S. Angelo, Edificio G, Via Cinthia, I-80126, Napoli, Italy.}

\author{Tiziana Di Salvo}
\email{tiziana.disalvo@unipa.it}
 \affiliation{Università degli Studi di Palermo, Dipartimento di Fisica e Chimica "Emilio Segré", Palermo, Italy.}
\author{Francesca Spinnato}
    \email{francesca.spinnato@community.unipa.it}
    \affiliation{Università degli Studi di Palermo, Dipartimento di Fisica e Chimica "Emilio Segré", Palermo,  Italy.}

%\date{\today} % Leave empty to omit a date

\begin{abstract}
We sketch  the main features of the Noether Symmetry Approach, a method to reduce and solve dynamics of physical systems by selecting Noether symmetries, which correspond to conserved quantities. Specifically, we take into account  the vanishing Lie derivative condition for general canonical Lagrangians to select symmetries. Furthermore, we extend the prescription to the first prolongation of the Noether vector. It is possible to show that the latter application provides a general constraint on the infinitesimal generator $\xi$, related to the spacetime translations. This approach can  be  used  for several applications. In the second part of the work,  we consider a gravity theory including the coupling between a  scalar field $\phi$ and the  Gauss-Bonnet topological  term $\mathcal{G}$. In particular, we study a gravitational action containing the function $F(\mathcal{G}, \phi)$ and select viable models by the existence of  symmetries. Finally, we evaluate the selected models in a spatially-flat cosmological background and use  symmetries to find exact solutions. 
\end{abstract}

\keywords{Noether symmetries; modified theories of gravity; exact solutions.}

\maketitle

\section{Introduction} \label{sec:introduction}
    Over the years, General Relativity (GR) has been confirmed  at various  energy and spacetime scales. It is capable of predicting with high accuracy phenomena in the weak  field limit  \cite{Misner:1973prb}, where the Newton theory of gravity failed,  including \emph{e.g.} the perihelion precession of Mercury orbit, the deflection of light by the Sun and the gravitational redshift of light rays. At the same time, it provided the final evidences of gravitational waves \cite{LIGOScientific:2016aoc} and black holes \cite{EventHorizonTelescope:2019dse, EventHorizonTelescope:2019pgp}. On the other hand, GR manifests shortcomings at UV and IR scales, suggesting that it is not the definitive theory of gravity \cite{Will:2014kxa, Nojiri:2017ncd, Odintsov:2023weg}. For example, the galaxy rotation curve \cite{ParticleDataGroup:2012pjm, Bosma:1981zz}, the current cosmic expansion of the Universe \cite{Frieman:2008sn, SupernovaSearchTeam:1998fmf}, the problem of singularities  \cite{Vachaspati:2006ki, Barcelo:2009tpa}, the unification between gravity and the other fundamental interactions \cite{Goroff:1985th, Birrell:1982ix, Weinberg:1988cp} represent shortcomings that, to be solved, need further ingredients out of the theory  or new conceptual approaches. Within the formalism of GR, the two former problems are addressed by considering dark matter and dark energy, respectively, which should account for the majority of the Universe content ($\sim$ 95\%) though they have never been directly detected. The latter issues is closely related to the lack of a self-consistent quantum gravity theory, because GR cannot be dealt under the same standard as the other interactions. Even considering a semi-classical approach, GR cannot be renormalized by means of the usual techniques, as incurable divergences occur at UV scales \cite{Niedermaier:2006wt, Percacci:2007sz, Bajardi:2021lwp}. In this framework, possible extensions/modifications of GR attempt to address these issues either by modifying the gravitational action or relaxing various assumptions of GR. To the latter category belong theories \emph{e.g.} with affine connections different than Levi-Civita \cite{Alexandrov:2002br, Maldacena:1997re, Nilles:1983ge, Rovelli:1997yv}, breaking the Lorentz invariance \cite{Horava:2009uw}, considering higher dimensions \cite{Rubakov:1983bb, Bajardi:2021hya, Qiang:2009fu, Rasouli:2022tmc}, \emph{etc.} (see \cite{Bajardi:2022ypn, Capozziello:2011et} for examples of  alternatives to GR). On the other hand, to the former category belong models including in the gravitational action functions of the scalar curvature \cite{Sotiriou:2008rp, Mishra:2018tqo}, higher-order curvature invariants \cite{Blazquez-Salcedo:2017txk, Stelle:1976gc}, coupling between geometry and dynamical scalar fields \cite{Halliwell:1986ja, Uzan:1999ch}. By relaxing the assumption of a gravitational Lagrangian linearly dependent on the scalar curvature, higher-order field equations occur in the metric formalism. Most often, extended theories of gravity are taken into account to address the late-time cosmic expansion of the Universe without introducing dark energy \cite{Capozziello:2002rd}. This is due to the fact that the right hand side of the gravitational field equations can be intended as an effective energy-momentum tensor given by geometry, mimicking dark energy as a curvature quintessence. 
    
    One of the most famous extended models is the so called $f(R)$ gravity, whose action contains a function of the scalar curvature. GR is restored as soon as $f(R) = R$. Another possibility is to consider functions of other higher-order invariants, such as $R^{\mu \nu \rho \sigma}R_{\mu \nu \rho \sigma}$ or $R^{\mu \nu}R_{\mu \nu}$, with $R^{\mu \nu \rho \sigma}$ and $R^{\mu \nu}$ being the Riemann and the Ricci tensors, respectively. Among all possible combinations given by these three curvature scalars, the only one providing a topological surface term is $\mathcal{G} = R^2 - 4 R^{\mu \nu}R_{\mu \nu}+ R^{\mu \nu \rho \sigma}R_{\mu \nu \rho \sigma}$, which is called \emph{Gauss-Bonnet} term. In four dimensions, the latter turns into the Euler density, meaning that once integrated over the manifold it provides the Euler characteristic. Being a topological surface, it does not contribute to the field equations (in four dimensions); however, a function of $\G$ is not trivial in four dimensions, then its contributions is relevant into dynamics. The relevance of dealing with the Gauss--Bonnet term is twofold: on the one hand, being a topological surface, it can contribute to the reduction of  the field equations, allowing to find analytic solutions; on the other hand, this term naturally emerges in gauge theories of gravity (such as Lovelock \cite{Cvetkovic:2016ios, Zanelli:2005sa} or Born-Infield \cite{Comelli:2005tn} gravity) and can be helpful to address issues emerging at UV scales. In particular, it can give ghost-free models.
    Most often, the action $S \sim \int (R + f(\G)) \, d^4 x$ is taken into account as a starting point, so that GR is safely recovered when $f(\G) \to 0$ and the latter can account for dynamical contributions to dark energy  \cite{Bajardi:2022tzn, Bajardi:2020osh}. 

In this paper, after reviewing the so-called {\bf Noether Symmetry Approach} \cite{cimento}, we will consider a gravitational action  with  non-minimal coupling between the Gauss--Bonnet term and a scalar field $\phi$, with corresponding kinetic and potential terms. 

The presence of the scalar field can be potentially useful to properly evaluate the very early times of the Universe evolution, when the inflationary epoch is supposed to be dominant. More precisely, we consider a function of both $\G$ and $\phi$, $F(\G, \phi)$, a kinetic term $\omega(\phi)$ and a potential term $V(\phi)$. This choice is due to the fact that, in some cosmological contexts, the Gauss--Bonnet term can act like the Ricci curvature according to the relation $\sqrt{|\G|} \sim R$, so that GR can be restored even without the presence of the scalar curvature. The unknown functions are thus derived by using  the Noether Symmetry Approach, a selection criterion aimed at finding models with symmetries \cite{Dialektopoulos:2018qoe, Bajardi:2022ypn, Urban:2020lfk}. In the majority of applications, the Noether Theorem is used to find out integrals of motion, arising from transformation laws that leave the action invariant. This procedure allows to reduce dynamics and, eventually, to integrate it. Here we show how to reverse the usual approach and select theories with symmetries starting from undefined actions. This point of view is taken into account in several works, as it represents a physical criterion capable to unveiling viable models among several  possible choices \cite{Capozziello:2007wc, Capozziello:1999xs, Basilakos:2011rx, Capozziello:1996ay}. The resulting constants of motion can be used to perform a proper change of variables (suggested by the Noether theorem) leading to a reduction of the minisuperspace dimension \cite{Capozziello:2012hm, Bajardi:2023vcc, Capozziello:2022vyd}. Reducing the dynamics of the system turns out to be physically relevant in any branch of physics, especially in those fields in which the equations of motion are difficult to be  handled, as well as theories of gravity. Here, by applying the Noether symmetry existence condition to the most general canonical field Lagrangian, we show that some constraints on the infinitesimal generator can be given from the beginning, so that the Noether system can be highly simplified. This is of extreme utility especially in those cases where the given model involves more curvature invariants and the application of the Noether system can result in many differential equations see \emph{e.g.} \cite{Bajardi:2022ypn, Bahamonde:2016grb}.
 
    The paper is organized as follows: in Sec. \ref{briefintro} we inroduce the Lie derivative and the Noether vector, also outlining their main features and possible applications. We thus apply the Noether symmetry approach to  general canonical Lagrangians; as a result, we obtain a general constraint on the symmetry generator, which is of particular interest in the application to modified theories of gravity. 
In Sec. \ref{quantizationby},  we outline the features of the Noether vector and its relation with the Lie derivative pointing out the role of Noether charges.   Realizations of the method are reported in Secs. \ref{XL}, \ref{XXL}, \ref{XXXL}.
    In Sec. \ref{sec:coupling} we adopt the Noether Symmetry Approach  to a scalar-tensor  action, containing the coupling between a dynamical scalar field and the Gauss--Bonnet term. We also find cosmological solutions for all selected models.  Conclusions are drawn in  Sec. \ref{sec:conclusions}.

    \section{The Noether Symmetry Approach: an overview} \label{briefintro}

Let us start by considering a  Noether vector $\displaystyle X = \alpha^i \frac{\partial}{\partial q^i} + \dot{\alpha}^i \frac{\partial}{\partial \dot{q}^i}$ and a generic Lagrangian $\Lagr$; let $L_X$ be the Lie derivative along the flux of the vector X. Here $q^i$ are the configuration space variables, namely the generalized coordinates, with the index $i$ running from $1$ to the number of variables of the configuration space; $\alpha^i$ are functions of the coordinates $q^i$ and the dot denotes the time derivative. The \textbf{Noether theorem} states that the condition $L_X \Lagr = 0$ implies that the phase flux is conserved along $X$ and a constant of motion exists. This is an alternative formulation of the theorem, which allows to suitably find integrals of motion by means of the vanishing Lie derivative condition. Specifically, if $L_X \Lagr = 0$, then the quantity 
\begin{equation}
\Sigma_0 \equiv \alpha^i \frac{\partial \Lagr}{\partial \dot{q}^i},
\label{sigma zero}
\end{equation}
is a constant of motion. The proof for this statement is straightforward and can be found \emph{e.g.} in \cite{Capozziello:1999xs, Bajardi:2022ypn}. Moreover, it can be shown that the above formulation of the Noether theorem is  a simplification involving only internal symmetries. To this purpose, let us consider a coordinate transformation of the form
\begin{equation}
\begin{cases}
\Lagr(t,q^i \dot{q}^i) \to \Lagr (\overline{t}, \overline{q}^i, \dot{\overline{q}}^i)
\\
\overline{t} = t + \epsilon \xi(t,q^i) + O(\epsilon^2)
\\
\overline{q}^i = q^i + \epsilon \alpha^i(t,q^i) + O(\epsilon^2),
\end{cases}
\label{trasformazione coordinate}
\end{equation} 
where $\{\overline{t}, \overline{q}^i\}$ are the variables of the transformed space-time, $\epsilon$ is an arbitrary constant, $\xi$ and $\alpha^i$ scalar and vector functions, respectively, of the coordinates.
The corresponding generator is
\begin{equation}
X^{[1]} = \xi \frac{\partial }{\partial t} + \alpha^i \frac{\partial }{\partial q^i} + \alpha^{i \; [1]} \frac{\partial}{\partial \dot{q}^i} = \xi \frac{\partial }{\partial t} + \alpha^i \frac{\partial }{\partial q^i} + (\dot{\alpha}^i - \dot{q}^i \dot{\xi}) \frac{\partial}{\partial \dot{q}^i}
\end{equation}
and it is the first prolongation of Noether vector. Moreover, assuming that the transformations \eqref{trasformazione coordinate} leave the Euler-Lagrange equations invariant, the identity
\begin{equation}
X^{[1]} \Lagr + \dot{\xi} \Lagr = \dot{g}(t,q^i) ,\label{Teorema}
\end{equation}
must hold; $g$ is an arbitrary function of coordinates and fields. It is called "gauge function". The resulting conserved quantity is a generalization of \eqref{sigma zero}:
\begin{equation}
I = \xi \left(\Lagr -  \dot{q}^i \frac{\partial \Lagr}{\partial \dot{q}^i}\right) + \alpha^i \frac{\partial \Lagr}{\partial \dot{q}^i} - g(t,q^i) .
\label{Conserved quantity}
\end{equation}
Eq. \eqref{Conserved quantity} turns into Eq. \eqref{sigma zero} when $\xi = 0$, that is when internal symmetries are considered.
The  demonstration of the above  Noether theorem can be found in \cite{Bajardi:2019zzs, Bajardi:2022ypn}.
In the next subsections we apply the Noether Theorem to the general  canonical Lagrangians, with the purpose to put general constraints on the symmetry generator. 

Let us then consider a canonical system with kinetic term
\begin{equation}
\mathcal{T} = \mathcal{A}_{ij}\left( q^i \right) \dot{q}^{i}\dot{q}^{j}
\end{equation}
and potential
\begin{equation}
\mathcal{U} = V\left( q^i\right) +A_{i}\left( q^i\right) \dot{q}^{i}.
\end{equation}
In the above equations, $\mathcal{A}_{ij}$ are matrix elements depending on the generalized coordinates $q^i$, $V(q^i)$ is a scalar function of the coordinates and $A_i$ is a vector function of $q^i$. Therefore, the Lagrangian $\Lagr = \mathcal{T} - \mathcal{U}$, can be written as
\begin{equation}
\mathcal{L}\left( q, \dot{q}\right) =\mathcal{A}_{ij}\left( q\right) \dot{q}^{i}\dot{q}^{j}-V\left( q\right) -A_{i}\left( q\right) \dot{q}^{i},
\label{lagran}
\end{equation}
which is the most general canonical Lagrangian depending on the coordinates $q^i$ and on their first derivatives. Note that, except for some particular cases, most modified gravity Lagrangians can be recast as  in Eq. \eqref{lagran}. For this reason, the constraint on the symmetry generator, provided in the next section, is of particular interest for several applications. It allows to constrain the space of solutions. In this way, the Noether system of differential equations can be reduced such that the approach can provide exact solutions. 

\subsection{ The Lie derivative and the Noether charges}\label{quantizationby}
Before considering some  applications of the above approach,  let us  discuss the general capability of Noether theorem to reduce the minisuperspace dimension, as well as the link between the condition \eqref{Teorema} and the Lie derivative. Let $GL(n,\mathbb{R})$ be the Lie group on  $\R$ consisting of a set of matrices $D_\R(\theta)$. Let us consider a generic transformation belonging to such a group.  For each transformation generator,  the corresponding  conserved quantities, ruled by the same algebra of the generator, are
\begin{equation}
\begin{cases}
\left[ X^i , X^j \right] = i f^{ij}_k  X^k
\\
\left\{\Sigma^i , \Sigma^j \right\} = i f^{ij}_k  \Sigma^k,
\end{cases}
\end{equation}
where $\Sigma^i$ are the integrals of motion, with the indexes $i,j,k$ running from 1 to the total number of symmetries, and $f^{ij}_k$ the structure constants of the given Lie algebra.  The Noether Theorem associates to any generator the corresponding conserved quantity, \emph{i.e.} the so called "Noether's current", which integrated over the hypersurface of the considered minisuperspace leads to the so called Noether's charge. Notice that when the Lagrangian depends on the spacetime variables $x^{\mu}$, the condition \eqref{Teorema} can be generalized as:
\begin{equation}
X^{[1]} \La + \partial_\mu \xi^\mu \La = \partial_\mu g^\mu,
\label{X1}
\end{equation}
with $\La$ being the Lagrangian density and $X^{[1]}$ the generalized first prolongation of Noether vector
\begin{equation}
X^{[1]} = \xi^\mu \partial_\mu + \alpha^i \frac{\partial }{\partial \phi^i} + (\partial_\mu \alpha^i - \partial_\mu \phi^i \partial_\nu \xi^\nu) \frac{\partial}{\partial (\partial_\mu \phi^i)}.
\end{equation}
Therefore, the application of \eqref{X1} to a given Lagrangian density $\La$, can be made explicit as:
\begin{eqnarray}
&& X^{[1]} \La + \partial_\mu \xi^\mu \La = \partial_\mu g^\mu  \to \nonumber
\\
&& \to \left\{ \xi^\mu \partial_\mu + \delta \phi^i \frac{\partial }{\partial \phi^i} + (\partial_\mu \delta \phi^i - \partial_\mu \phi^i \partial_\nu \xi^\nu) \frac{\partial}{\partial (\partial_\mu \phi^i)} + \partial_\mu \xi^\mu \right\} \La = \partial_\mu g^\mu \to \nonumber
\\
&& \to \left\{ L_X - \partial_\nu \xi^\nu \left(\partial_\mu \phi^i \frac{\partial}{\partial (\partial_\mu \phi^i)} - 1 \right) \right\} \La = \partial_\mu g^\mu \to \nonumber
\\
&& \to L_X \La - \partial_\nu \xi^\nu {\mathscr{T}} = \partial_\mu g^\mu,
\label{lie lagr}
\end{eqnarray}
where ${\mathscr{T}} $ is the trace of the energy-momentum tensor. For  time-depending fields, the condition \eqref{lie lagr}  becomes:
\begin{equation}
L_X \Lagr - \dot{\xi} \mathcal{H} = \dot{g}\,,
\label{relaz lie extedned}
\end{equation}
where $\mathcal{H}$ is the associated Hamiltonian, as we will see the example below.
Eq. \eqref{relaz lie extedned} generalizes the condition of vanishing Lie derivative along the flux of $X$ to its first prolongation $X^{[1]}$. 
 
In addition, besides finding the symmetries of the given Lagrangians, the Lie derivative can be also used to find the symmetries of the metric tensor. Indeed, the application of the operator $L$ along the flux of a given vector field $X$ to the metric tensor, yields:
\begin{equation}
L_X g_{\mu \nu} = 2 g_{\mu \nu}  \phi(x^\mu),
\end{equation}
from which the conformal Killing vectors can be split in three categories: 
\begin{itemize}
\item \emph{Proper}  Killing  Vector\index{Killing Vector} $\, \to \, \phi(x^\mu) \neq 0 \, \to \,  L_X g_{\mu \nu} = 2 g_{\mu \nu}  \phi(x^\mu)$
\item \emph{Special} Killing Vector\index{Killing Vector} $\, \to \, \phi(x^\mu) = 0 \, \to \, L_X g_{\mu \nu} = 0$
\item \emph{Homotetic} Killing Vector\index{Killing Vector} $\, \to \, \partial_\mu \phi(x^\mu) = 0 \, \to \, L_X g_{\mu \nu} \sim$ const .
\end{itemize}
In particular, for Special Killing vectors, the well known Killing equation
\begin{equation}
L_\xi g_{\mu \nu} = [\xi, g_{\mu \nu}] = 0 \;\; \to \;\; D_\nu \xi_{\mu} - D_\mu \xi_\nu  = 0,
\end{equation}
for the isometries automatically follows. With these considerations in mind, let us now develop some specific examples.

\subsection{Noether Symmetries in the Canonical Two-Particle Lagrangian}\label{XL}

Starting from Eq. \eqref{lagran}, let us  apply the Noether Symmetry Approach to a two-particle Lagrangian
\begin{equation}
\begin{split}
 \mathcal{L}\left( q_{1}, q_{2}, \dot{q}_{1}, \dot{q}_{2} \right) =&\mathcal{A}_{11}\left( q_{1}, q_{2}\right) {\dot{q}_{1}}^{~2}+\mathcal{A}_{12}\left( q_{1}, q_{2}\right) \dot{q}_{1}\dot{q}_{2}+\mathcal{A}_{21}\left( q_{1}, q_{2}\right) \dot{q}_{1}\dot{q}_{2}+\\+&\mathcal{A}_{22}\left( q_{1}, q_{2}\right) {\dot{q}_{2}}^{~2}-V\left( q_{1}, q_{2}\right)-A_{1}\left( q_{1}, q_{2}\right)\dot{q}_{1}-A_{2}\left( q_{1}, q_{2}\right)\dot{q}_{2}.
 \end{split}
 \label{Lagra12}
\end{equation}
The search for internal symmetries can be pursued by setting the Lie derivative of $\Lagr$ equal to zero. The Noether vector,  corresponding to a general transformation of  variables $q^i$, reads as:
\begin{equation}
X=\alpha\left( q_{1}, q_{2}\right)\partial_{q_{1}}+\beta\left( q_{1}, q_{2}\right)\partial_{q_{2}}+\dot{\alpha}\left( q_{1}, q_{2}\right)\partial_{\dot{q}_{1}}+\dot{\beta}\left( q_{1}, q_{2}\right)\partial_{\dot{q}_{2}},
\end{equation}
where $\alpha$ ($\beta$) is a generic function of the two variables $q_1$ and $q_2$, and $\dot{\alpha}$ ($\dot{\beta}$) its total time derivative, namely $\dot{q}_1 \partial_{q_1} \alpha + \dot{q}_2 \partial_{q_2} \alpha$, with $\partial_{q_i}$ denoting the derivative with respect to $q_i$.
The condition $ L_{X}\mathcal{L}=0$, asking for the cancellation of derivative terms, leads to the following system of differential equations
\begin{subequations}
\begin{align}
&\alpha\partial_{q_{1}}V+\beta\partial_{q_{2}}V=0\\
\dot{q}_{1}:~\quad& \alpha\partial_{q_{1}}A_{1}+\beta\partial_{q_{2}}A_{1}+A_{1}\partial_{q_{1}}\alpha+A_{2}\partial_{q_{1}}\beta=0\\
\dot{q}_{2}:~\quad& \alpha\partial_{q_{1}}A_{2}+\beta\partial_{q_{2}}A_{2}+A_{1}\partial_{q_{2}}\alpha+A_{2}\partial_{q_{2}}\beta=0\\
{\dot{q}_{1}}^{~2}:~\quad& \alpha\partial_{q_{1}}\mathcal{A}_{11}+\beta\partial_{q_{2}}\mathcal{A}_{11}+2\left( \partial_{q_{1}}\alpha\right) \mathcal{A}_{11}+\left( \partial_{q_{1}}\beta\right) \mathcal{A}_{12}+\left( \partial_{q_{1}}\beta\right) \mathcal{A}_{21}=0\\
{\dot{q}_{2}}^{~2}:~\quad& \alpha\partial_{q_{1}}\mathcal{A}_{22}+\beta\partial_{q_{2}}\mathcal{A}_{22}+\left( \partial_{q_{2}}\alpha\right) \mathcal{A}_{12}+\left( \partial_{q_{2}}\alpha\right) \mathcal{A}_{12}+2\left( \partial_{q_{2}}\beta\right) \mathcal{A}_{22}=0\\
\dot{q}_{1}\dot{q}_{2}:\quad 
\begin{split}
&\alpha\partial_{q_{1}}\mathcal{A}_{12}+\alpha\partial_{q_{1}}\mathcal{A}_{21}+\beta\partial_{q_{2}}\mathcal{A}_{12}+\beta\partial_{q_{2}}\mathcal{A}_{21}+\\&+2\left( \partial_{q_{2}}\alpha\right) \mathcal{A}_{11}+\left( \partial_{q_{1}}\alpha\right) \mathcal{A}_{12}+\left( \partial_{q_{1}}\alpha\right) \mathcal{A}_{21}+\\&+\left( \partial_{q_{2}}\beta\right) \mathcal{A}_{12}+\left( \partial_{q_{2}}\beta\right) \mathcal{A}_{21}+2\left( \partial_{q_{1}}\beta\right) \mathcal{A}_{22}=0.
\end{split}
\label{sistema0}
\end{align}
\end{subequations}
According to the same procedure, it is possible to get a system of differential equations coming from the application of the identity \eqref{Teorema} to the Lagrangian \eqref{Lagra12}. Specifically, being the condition $L_X \Lagr = 0$ contained in the general identity \eqref{Teorema}, the system related to the latter will be contained in  the former one. As a matter of fact, the system involving also external transformations reads:
\begin{subequations}
	\begin{align}
	&\alpha\partial_{q_{1}}V+\beta\partial_{q_{2}}V+\left( \partial_{t}\alpha\right) A_{1}+\left( \partial_{t}\beta\right) A_{2}+\left( \partial_{t}\xi\right) V-\partial_{t}g=0\\
	\dot{q}_{1}:~\quad
	\begin{split}
	&-\alpha\partial_{q_{1}}A_{1}-\beta\partial_{q_{2}}A_{1}-A_{1}\partial_{q_{1}}\alpha+2\left( \partial_{t}\alpha\right) \mathcal{A}_{12}-A_{2}\partial_{q_{1}}\beta+\\&+\left( \partial_{t}\beta\right)\mathcal{A}_{12}+\left( \partial_{t}\beta\right)\mathcal{A}_{21}-\partial_{q_{1}}g =0
	\end{split}\\
	\dot{q}_{2}:~\quad
	\begin{split}
	&-\alpha\partial_{q_{1}}A_{2}-\beta\partial_{q_{2}}A_{2}-A_{1}\partial_{q_{2}}\alpha+\left( \partial_{t}\alpha\right) \mathcal{A}_{12}+\left( \partial_{t}\alpha\right) \mathcal{A}_{21}-\\&-A_{2}\partial_{q_{2}}\beta+2\left( \partial_{t}\beta\right)\mathcal{A}_{22}-\partial_{q_{2}}g =0
	\end{split}\\
	{\dot{q}_{1}}^{~2}:~\quad
	\begin{split}
	&\alpha\partial_{q_{1}}\mathcal{A}_{11}+\beta\partial_{q_{2}}\mathcal{A}_{11}+2\left( \partial_{q_{1}}\alpha\right) \mathcal{A}_{11}\\&-\left( \partial_{t}\xi\right) \mathcal{A}_{11}+\left( \partial_{q_{1}}\beta\right)\mathcal{A}_{12}+\left( \partial_{q_{1}}\beta\right)\mathcal{A}_{21} =0
	\end{split}\\
	{\dot{q}_{2}}^{~2}:~\quad
	\begin{split}
	&\alpha\partial_{q_{1}}\mathcal{A}_{22}+\beta\partial_{q_{2}}\mathcal{A}_{22}+\left( \partial_{q_{2}}\alpha\right) \mathcal{A}_{12}+\\&+\left( \partial_{q_{2}}\alpha\right) \mathcal{A}_{21}+2\left( \partial_{q_{2}}\beta\right) \mathcal{A}_{22}-\left( \partial_{q}\xi\right)\mathcal{A}_{22}=0
	\end{split}\\
	\dot{q}_{1}\dot{q}_{2}:\quad 
	\begin{split}
	&\alpha\partial_{q_{1}}\mathcal{A}_{12}+\alpha\partial_{q_{1}}\mathcal{A}_{21}+\beta\partial_{q_{2}}\mathcal{A}_{12}+\beta\partial_{q_{2}}\mathcal{A}_{21}+\\&+2\left( \partial_{q_{2}}\alpha\right) \mathcal{A}_{11}+\left( \partial_{q_{1}}\alpha\right) \mathcal{A}_{12}+\left( \partial_{q_{1}}\alpha\right) \mathcal{A}_{21}-\left( \partial_{t}\xi\right) \mathcal{A}_{12}-\\&-\left( \partial_{t}\xi\right) \mathcal{A}_{21}+\left( \partial_{q_{2}}\beta\right) \mathcal{A}_{12}+\left( \partial_{q_{2}}\beta\right) \mathcal{A}_{21}+2\left( \partial_{q_{1}}\beta\right) \mathcal{A}_{22}=0
\end{split}	\\
\dot{q}_{1}^{~2}\dot{q}_{2}:~\quad &-\left( \partial_{q_{1}}\xi\right) \mathcal{A}_{12}-\left( \partial_{q_{1}}\xi\right) \mathcal{A}_{21}-\left( \partial_{q_{2}}\xi\right) \mathcal{A}_{11}=0	
\\
\dot{q}_{1}\dot{q}_{2}^{~2}:~\quad &-\left( \partial_{q_{2}}\xi\right) \mathcal{A}_{12}-\left( \partial_{q_{2}}\xi\right) \mathcal{A}_{21}-\left( \partial_{q_{1}}\xi\right) \mathcal{A}_{22}=0	
\\\dot{q}_{1}^{~3}:~\quad &-2\left( \partial_{q_{1}}\xi\right) \mathcal{A}_{11}+\left( \partial_{q_{1}}\xi\right) \mathcal{A}_{11}=0	
\\
\dot{q}_{2}^{~3}:~\quad &-2\left( \partial_{q_{2}}\xi\right) \mathcal{A}_{22}+\left( \partial_{q_{2}}\xi\right) \mathcal{A}_{22}=0,
	\label{sistema1}
	\end{align}
\end{subequations}
which is a generalization of the above one. Note that $\mathcal{A}_{ij}$ are arbitrary functions, so that the only possible solution of the last two equations is $\xi \equiv \xi(t)$, which means that the infinitesimal generator $\xi$ must depend on time only. In view of this result, we can neglect all  terms containing other partial derivatives of $\xi$, as well as $\partial_{q_{1}}\xi$ or $\partial_{q_{2}}\xi$. The function $\xi(t)$ plays an important role in the Noether approach, since it gives rise to  the spacetime transformations. Once we set $\xi= 0$, the generator $X^{[1]}$ turns into $X$ and the identity \eqref{Teorema} folds into the condition of vanishing Lie derivative. This relation can be explicitly obtained by considering the application of the Lie derivative, namely
\begin{equation}
L_X \Lagr = X \Lagr =\left( \alpha^i \partial_{q^i} + \dot{\alpha}^i  \partial_{\dot{q}^i} \right) \Lagr. \label{lie1}
\end{equation}
By merging Eq. \eqref{Teorema} and Eq. \eqref{lie1}, we obtain
\begin{equation}
L_X \Lagr + \dot{\xi} {\cal{H}} = \dot{g},
\end{equation}
being ${\cal{H}}$ the Hamiltonian function, defined as
\begin{equation}
\mathcal{H} = \dot{q}^i \frac{\partial \Lagr}{\partial \dot{q}^i} - \Lagr.
\end{equation}
Considering that the gauge field $g$ can be arbitrarily set to zero, the above result shows that the imposition $\dot{\xi} \sim const$ implies
\begin{equation}
L_X \Lagr = \xi_0 \cal{H} .
\end{equation}
Notice that, in this example, we only considered time-depending Lagrangians, though the method can be even applied to general systems depending on other spatial coordinates \cite{Bajardi:2019zzs, Bahamonde:2019swy, Bahamonde:2019jkf}. In the next subsections, we analyze canonical Lagrangians of different form, finding out the corresponding symmetry generators and conserved quantities.

\subsection{Example I:  The free particle}\label{XXL}
The simplest case is the free particle Lagrangian, which can be obtained from Eq. \eqref{lagran} by setting $V(q)=A(q)=0$. Moreover, for the sake of simplicity, we also set $\mathcal{A}_{ij}=\delta_{ij}$. The Lagrangian therefore takes the form
\begin{equation}
\mathcal{L}= (\dot{q})^2.
\end{equation}
The application of the vanishing Lie derivative condition to the above Lagrangian yields the only equation
\begin{equation}
\frac{\partial \alpha}{\partial q} = 0,
\end{equation}
which automatically leads to $\alpha = \alpha_0$, with $\alpha_0$ constant. The system is invariant under local translation whose symmetry generator is $X = \alpha_0 \partial_q$. The conserved quantity can be found as a subcase of Eq. \eqref{Conserved quantity}, which provides
\begin{equation}
\Sigma_0 = \alpha \frac{\partial \mathcal{L}}{\partial \dot{q}} = \alpha_0 \frac{\partial \mathcal{L}}{\partial \dot{q}} = \alpha_0 \pi_q,
\end{equation}
namely the conjugate momentum, as expected.

\subsection{Example II: The harmonic oscillator}\label{XXXL}
Let us apply the Noether Symmetry Approach to a Lagrangian with harmonic potential, considering only the vanishing Lie derivative condition. Starting  from the above general Lagrangian, it  contains only one set of generalized coordinates $q^i$. Thus the two methods turns out to be  equivalent. 

We expect the angular momentum to be the conserved quantity provided by the Noether theorem. The canonical form for the harmonic potential and the kinetic term can be obtained by setting $\mathcal{A}_{11}=\mathcal{A}_{22}=1$, $\mathcal{A}_{12}=\mathcal{A}_{21}=A_{1}=A_{2}=0$ and $V=q_{1}^{~2}+q_{2}^{~2}$, so that  Lagrangian \eqref{Lagra12} becomes
\begin{equation}
\mathcal{L}=\dot{q}_{1}^{~2}+\dot{q}_{2}^{~2}-q_{1}^{~2}-q_{2}^{~2} .
\label{lagra}
\end{equation}
The system (11) reduces to four differential equations:
\begin{subequations}
\begin{align}
&2\alpha q_{1}+2\beta q_{2}=0\\
&2\partial_{q_{1}}\alpha=0\\
&2\partial_{q_{2}}\beta=0\\
&2\partial_{q_{2}}\alpha+2\partial_{q_{1}}\beta=0.
\label{sistema oscillatore}
\end{align}
\end{subequations}
whose solution provides the following symmetry generator
\begin{equation}
X=q_{2}\partial_{q_{1}}-q_{1}\partial_{q_{2}},
\end{equation}
which is nothing but the $SO(2)$ group generator, as expected. Similarly, the conserved quantity $\Sigma_0$ can be found by means of the condition $\displaystyle \Sigma_{0}=\alpha^{i}\dfrac{\partial\mathcal{L}}{\partial \dot{q}^{i}}$, that is
\begin{equation}
\Sigma_{0}=2q_{2}\dot{q_{1}}-2q_{1}\dot{q_{2}},
\end{equation}
which is the orbital angular momentum.

    \section{An application: Coupling Gauss-Bonnet gravity to a scalar field} \label{sec:coupling}
   Let us now apply the prescription of the previous section to a modified gravity Lagrangian including the coupling between  the Gauss--Bonnet term $\G$ and a scalar field $\phi$ through the coupling function $F(\G, \phi)$, the kinetic and the potential terms (namely $\omega(\phi)$ and $V(\phi)$, respectively). It reads as:
\begin{equation}
    S = \int \sqrt{-g} \biggl[ F(\phi, \mathcal{G}) + \omega(\phi)\partial_{\mu}\phi\partial^{\mu}\phi + V(\phi)\biggr]d^{4}x,
\end{equation}
with $g$ being the determinant of the metric and with $\mu$ running from 0 to 4. We focus on a Friedman-Robertson-Walker (FRW) spatially-flat line element of the form
\begin{equation}
    \nonumber ds^2 = dt^2 - a^2(t) \delta_{ij} dx^i dx^j,
\end{equation}
where $\delta_{ij}$ is the three-dimensional unitary matrix, $x^i$ the spatial coordinates and $a(t)$ the scale factor, depending on the cosmic time $t$ only. By this choice, the Gauss--Bonnet term becomes
\begin{equation}
    \G = 24 \frac{\dot{a}^2 \ddot{a}}{a^3} = \frac{8}{a^3} \frac{d}{dt}\left( \dot{a}^3\right)
    \label{exprG}
\end{equation}
and can be adopted to obtain the point-like Lagrangian. To this purpose, let us consider the Lagrange multipliers method with constraint given by Eq. \eqref{exprG} and integrate the three-dimensional hypersurface, so that the action can be recast as:
\begin{equation}
    S = \int a^3 \left[F(\G, \phi) - \lambda \left( \G - \frac{8}{a^3} \frac{d \dot{a}^3}{dt}\right) + \omega(\phi) \dot{\phi}^2 + V(\phi) \right] dt,
    \label{pointaction}
\end{equation}
with $\lambda$ being the Lagrange multiplier. Notice that, with the above ansatz for the line element, a function of the Gauss-Bonnet term can behave like the Ricci scalar curvature, as soon as $f(\G) \sim \sqrt{\G}$, due to the reasons mentioned in the introduction. Therefore, in cosmological space-times, the most known scalar-tensor actions can be recovered by proper choices of the coupling function. For instance, the Brans-Dicke model can be obtained by setting $F(\phi, \G) \sim \phi \sqrt{\G}\,$, $\omega(\phi) \sim 1/\phi$ and $V(\phi) = 0$. This is worth to be mentioned since, as showed in Fig. 1, most symmetries can yield the Brans-Dicke action by setting $w = 1, k =1/2, q=-1$ and $V_0 = 0$.

By varying the action with respect to $\G$ and equating the result to zero, we straightforwardly find $ \lambda = F_\G(\G, \phi)$, where the subscript denotes the derivative with respect to the Gauss--Bonnet term. Now, integrating by parts the second derivatives occurring in Eq. \eqref{pointaction} and neglecting the boundaries, the point-like Lagrangian can be written as:
\begin{equation}
  \mathcal{L} =  a^{3} \biggl (F- \dfrac{\partial F}{\partial \mathcal{G}}\mathcal{G} \biggr) +  a^{3} \omega(\phi) \dot{\phi}^{2} + a^3 V(\phi) - 8\dot{a}^{3} \biggl(\dfrac{\partial^{2}F}{\partial \mathcal{G}^{2}} \dot{\mathcal{G}} + \dfrac{\partial^{2}F}{\partial \phi \partial \mathcal{G}} \dot{\phi} \biggr).
  \label{initiallagr}
\end{equation}
In the three-dimensional minisuperspace $\mathcal{S} \sim \{a, \phi, \G\}$, there are three Euler--Lagrange equations that, together with the energy condition, provide the following dynamical system:
\begin{equation}
    \begin{cases}
        8 \dot{a}^2 \biggl(\ddot{\mathcal{G}}\dfrac{\partial^{2}F}{\partial\mathcal{G}^{2}}+2 \dot{\mathcal{G}} \dot{\phi} \dfrac{\partial^{3}F}{\partial \phi \partial\mathcal{G}^{2}}+ \dot{\mathcal{G}}^{2} \dfrac{\partial^{3}F }{\partial \mathcal{G}^{3}} +\ddot{\phi} \dfrac{\partial^{2}F}{\partial \phi \partial \mathcal{G}} + \dot{\phi}^{2} \dfrac{\partial^{3} F}{\partial \phi^{2} \partial \mathcal{G}}\biggr)+16\dot{a}\ddot{a}\biggl(\dot{\mathcal{G}} \dfrac{\partial^{2}F}{\partial \mathcal{G}^{2}} +\dot{\phi} \dfrac{\partial^{2}F}{\partial \phi \partial \mathcal{G}} \biggr)\\+ a^{2} \biggl(F - \mathcal{G} \dfrac{\partial F}{\partial \mathcal{G}}+V(\phi)+\omega(\phi)\dot{\phi}^{2} \biggr) =0 \\\\
        -24 \dot{a}^2 \ddot{a} \dfrac{\partial^{2}F}{\partial \phi \partial \mathcal{G}}+a^3 \biggl(-\dfrac{\partial F}{\partial \phi}+\mathcal{G}\dfrac{\partial^{2}F}{\partial \phi \partial \mathcal{G}}-\dfrac{\partial V (\phi)}{\partial \phi}+2 \omega (\phi) \ddot{\phi}+\dot{\phi}^2 \dfrac{\partial \omega (\phi)}{\partial \phi} \biggr)+6 a^2 \dot{a} \omega (\phi) \dot{\phi} = 0\\\\
          \mathcal{G} = 24 \dfrac{\dot{a}^{2}\ddot{a}}{a^{3}}\\\\
        24 \dot{a}^{3} \biggl(\dot{\mathcal{G}} \dfrac{\partial^{2} F}{\partial \mathcal{G}^{2}}+\dot{\phi} \dfrac{\partial^{2} F}{\partial \phi \partial \mathcal{G}} \biggr)+a^3 \biggl(F-\mathcal{G} \dfrac{\partial F}{\partial \mathcal{G}}+V(\phi)-\omega(\phi) \dot{\phi}^{2}\biggr) =0.
    \end{cases}
\end{equation}
The first two equations are the Euler-Lagrange equations with respect to the cosmological scale factor and the scalar field, respectively. Specifically, the second equation corresponds to the Klein-Gordon equation coming from the variation of the action with respect to the scalar field. The third equation is the one related to the Gauss--Bonnet term given by the Lagrange multiplier. Finally, the last equation is the energy condition, namely $E_\Lagr = \dot{q}^i \partial \Lagr/\partial \dot{q}^i - \Lagr = 0$, where $q^i$ are the generalized coordinates of the considered minisuperspace.
In order for the Lagrangian approach to be equivalent to the variational approach, the latter equation must be included to recover the "00" component of the  field equations. Specifically, the zero energy condition accounts for a further constraint completing the dynamical system, physically implying the Hamiltonian is zero on the constraint surface due to time reparameterization invariance \cite{Agrawal:2020xek}. As a result, the variational approach, naturally providing two different Friedman equations by the variation with respect to the metric, turns out to be completely equivalent to the Lagrangian approach.

Before applying the Noether symmetry prescription to the above Lagrangian, let us point out that the minisuperspace is three-dimensional and therefore the first prolongation of the Noether vector in this case reads:
\begin{equation}
    X^{[1]} = \xi \partial_t + \alpha \partial_a + \beta \partial_\G + \gamma \partial_\phi + (\dot{\alpha} - \dot{\xi} \dot{a}) \partial_{\dot{a}} + (\dot{\beta} - \dot{\xi} \dot{\G}) \partial_{\dot{\G}} + (\dot{\gamma} - \dot{\xi} \dot{\phi}) \partial_{\dot{\phi}} 
\end{equation}
where $\alpha, \beta, \gamma$ and $\xi$ are functions of $a, \G, \phi, t$, with $t$ being the cosmic time. By applying the Noether identity \eqref{Teorema} to the Lagrangian \eqref{initiallagr}, we get a system of differential equations that can be  slightly simplified by the imposition  $\xi = \xi(t)$. According to the discussion of the previous section, this can be assumed \emph{a priori} for canonical point-like Lagrangians. Thanks to this \emph{ansatz}, the system can be solved analytically, providing the explicit expressions of $F, \omega$ and $V$. Eighteen possible solutions are summarized in Fig. 1 ($f_0, \omega_0, V_0, k, w, p, q$ are real constants).
\begin{figure}[h!]
\centering
\includegraphics[width=0.6\linewidth]{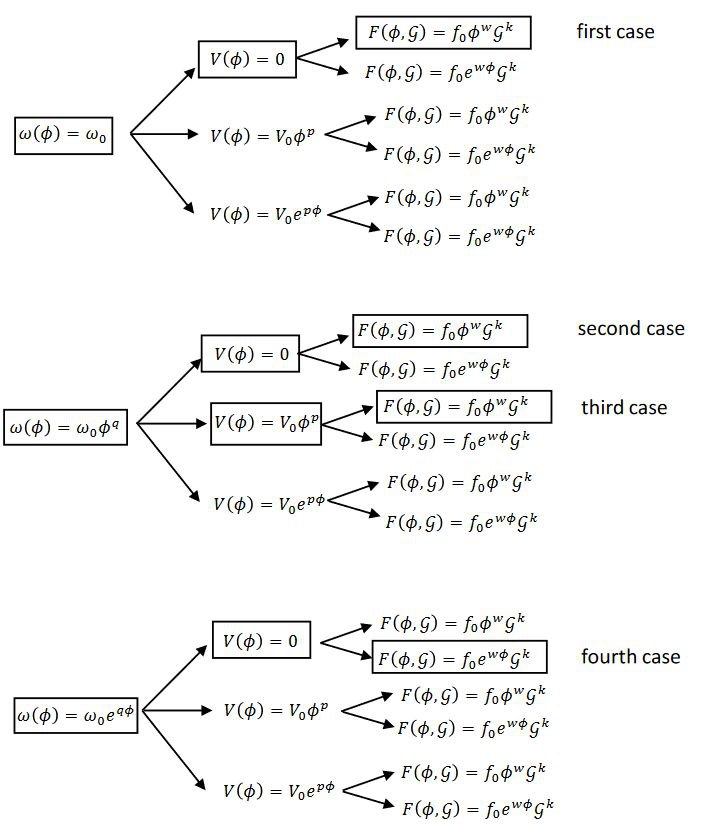}
\caption{Possible forms of the Lagrangian function selected by the existence of symmetries.}
\end{figure}
\newpage
For cosmological purposes, here we focus on the most interesting cases, which in Fig. 1 are marked by a rectangle, since they are the only symmetries leading either to time power-law or exponential de Sitter-like solutions. The other symmetries yield more complicated cosmological solutions, or equations of motions not solvable analytically. Though the latter category is still worth to be analysed, the study of numeric solutions is not the main goal of this paper. In the next subsections we analyse each case and find the corresponding cosmological solutions to the Euler--Lagrange equations.
\subsection{First case}
The first symmetry of the Lagrangian \eqref{initiallagr} occurs by setting $\omega(\phi) \sim \omega_0$ and $V(\phi) = 0$, with $\omega_0$ being constant. In this case, the point-like Lagrangian reduces to:
\begin{equation}
  \mathcal{L} = a^{3}\omega_{0} \dot{\phi}^{2}-f_{0}(k-1)\mathcal{G}^k \phi^{w}a^{3}-8f_{0}k \dot{a}^{3} \mathcal{G}^{k-2} \phi^{w-1} [(k-1) \phi \dot{\mathcal{G}}+w \mathcal{G} \dot{\phi}].
\label{eq:point_lagrangian1}
\end{equation}
with infinitesimal symmetry generators of the form
\begin{equation}
\begin{split}
    &\alpha(a) = \alpha_{0}a, \hspace{0.3cm} \beta(\mathcal{G}) = -4\xi_{0}\mathcal{G}, \hspace{0.3cm} \gamma(\phi) = \dfrac{6(1-2k)\alpha_{0}}{-2+8k-w} \phi, \hspace{0.3cm} \xi(t) = -\dfrac{3(w-2)\alpha_{0}}{-2+8k-w}t,
    \label{eq:sol1}
\end{split}
\end{equation}
where $\alpha(a), \beta(\mathcal{G}), \gamma(\phi)$ and $\xi(t)$ are the infinitesimal generators of the coordinate transformations. Notice that, in agreement with Eq. \eqref{sistema1}, the infinitesimal generator $\xi$ is a function of the sole time. In order to find analytic solutions to the equations of motion related to the Lagrangian \eqref{eq:point_lagrangian1}, we can set $w = 2$, so that the coupling function becomes $F(\phi, \mathcal{G}) = f_{0}\phi^{2}\mathcal{G}^{k}$, meaning that the general gravitational action is:
\begin{equation}
    S = \int \sqrt{-g} \left[f_0 \phi^{2}\mathcal{G}^{k} + \omega_0 \partial^\mu \phi \partial_\mu \phi \right] d^4 x.
\end{equation}
By these \emph{ansatz}, the explicit forms of the scale factor and the scalar field are
\begin{equation}
    a(t) = a_{0}e^{mt}, \hspace{0.3 cm} \phi(t) = \phi_{0}e^{nt}, \hspace{0.3 cm} \G = 24 m^4,
\end{equation}
namely de Sitter-like solutions. Here $a_{0}, m, \phi_{0}, n$ are real constants. A further constraint arising from the solution of the Euler--Lagrange equations concerns the relation among $\omega_0, f_0, n$ and $m$. Specifically, these four constants must satisfy the following system of algebraic equations:
\begin{equation}
    \begin{cases}
        f_{0} (24m^{4})^k [3 (k-1) m^2-4 k m n-4 k n^2]-3 m^2 n^2 \omega_{0} = 0\\
        f_{0} (24m^{4})^k-n \omega_{0}  (3 m+n)=0\\
        f_{0} (24m^{4})^k [(k-1) m-2 k n]+m n^2 \omega_{0}=0.
    \end{cases}
\end{equation}
Solving the third equation with respect to $n$, we obtain two solutions. After choosing one, we can solve the energy equation with respect to the constant $f_{0}$ to obtain two other solutions. As a result, we have a total of four possible combinations, but only three of these are valid.
For example, let us consider the cases $k=1$ and $k=1/2$; in particular, the latter is the value of $k$ allowing to recover GR at cosmological scales \cite{Bajardi:2020osh}. For these two values, we get the solutions:
\begin{equation}
k=1/2, \hspace{1cm} n = -\dfrac{3}{2}m, \hspace{1cm} f_{0} = -\dfrac{3\sqrt{6}}{16}\omega_{0},
\end{equation}
\begin{equation}
    k=1, \hspace{1cm} n = -\dfrac{5}{2}m, \hspace{1cm} f_{0} = -\dfrac{5\omega_{0}}{96m^{2}}.
\end{equation}

%Solving the third equation with respect to $n$, we obtain two solutions. After choosing one, we solve the energy equation with respect to the constant $f_{0}$. We again obtain two solutions. As a result, we have a total of four possible combinations of solutions, but only three of these are valid. 
\subsection{Second case}
Another symmetry leading to exact cosmological solutions occurs once considering a vanishing potential, a power-law form of the kinetic term $\omega(\phi)$, \textit{i.e.} $\omega(\phi) = \omega_{0}\phi^{q}$, and a coupling function of the form $F(\phi, \mathcal{G}) = f_{0}\phi^{w}\mathcal{G}^{k}$. The Lagrangian thus reads:
\begin{equation}
\mathcal{L} = a^{3}[\omega_{0}\phi^{q} \dot{\phi}^{2}-f_{0}(k-1)\mathcal{G}^k \phi^{w} ]-8f_{0}k \dot{a}^{3} \mathcal{G}^{k-2} \phi^{w-1} [(k-1) \phi \dot{\mathcal{G}}+w \mathcal{G} \dot{\phi}].
\end{equation}

The existence of the Noether symmetry selects the following infinitesimal generators:
\begin{equation}
    \alpha(a) = \alpha_{0}a, \hspace{0.3cm} \beta(\mathcal{G}) = -4\xi_{0}\mathcal{G}, \hspace{0.3cm} \gamma(\phi) = \dfrac{6(1-2k)\alpha_{0}}{(2+q)(4k-1)-w} \phi, \hspace{0.3cm} \xi(t) = -\dfrac{3(2+q-w)\alpha_{0}}{(2+q)(4k-1)-w}t,
\label{eq:sol2}
\end{equation}
and, as before,  the infinitesimal generator $\xi$ is a function of the sole time. The system of Euler-Lagrange equations, together with the energy conditions, can be solved  by setting $w = 2+q$ and provides the following exponential solution:
\begin{equation}
    a(t) = a_{0}e^{mt}, \hspace{0.3 cm} \phi(t) = \phi_{0}e^{nt},  \hspace{0.3 cm} \G = 24 m^4
\end{equation}
where $n$ and $m$ are real constants. Due to the constraint imposed by the equations of motion, the infinitesimal generators become
\begin{equation}
    \alpha(a) = \alpha_{0}a, \hspace{0.3cm} \beta(\mathcal{G}) = 0, \hspace{0.3cm} \gamma(\phi) = -\dfrac{3\alpha_{0}}{q+2} \phi, \hspace{0.3cm} \xi(t) = 0
\end{equation}
and the modified scalar-tensor action results 
\begin{equation}
    S = \int \sqrt{-g} \left[f_0 \phi^{q+2} \G^k + \omega_0 \phi^q \partial_\mu \phi \partial^\mu \phi \right] d^4x.
\end{equation}
Also here, there is a further constraint coming from the equations of motion which sets the relation among the constants $f_0, \omega_0, m, n$, namely:
\begin{equation}
    \begin{cases}
        f_{0} (24m^4)^k[3 (k-1) m^2-2 k m n (q+2)-k n^2 (q+2)^2]-3 m^2 n^2 \omega_{0}= 0\\
        n \omega_{0} [6 m+n (q+2)] -f_{0} (q+2) (24m^4)^k=0\\
        f_{0} (24m^4)^k [(k-1) m-k n (q+2)]+m n^2 \omega_{0} = 0.
    \end{cases}
\end{equation}
As the previous case, the solution of the above system results in four possible combinations for $f_{0}$ and $n$, but again only three are cosmologically meaningful.
Considering the same values of $k$ as before, we obtain:
\begin{equation}
k=1/2, \hspace{1cm} n = -\dfrac{3}{q+2}m, \hspace{1cm} f_{0} = -\dfrac{3\sqrt{6}}{4(q+2)^{2}}\omega_{0},
\end{equation}
\begin{equation}
    k=1, \hspace{1cm} n = -\dfrac{5}{2+q}m, \hspace{1cm} f_{0} = -\dfrac{5\omega_{0}}{24(q+2)^{2}m^{2}}.
\end{equation}

\subsection{Third case}
The third symmetry arising from the solution of the Noether system  involves both the coupling function, the kinetic and the potential terms. Specifically, the Noether system can be solve if $\omega(\phi) = \omega_{0}\phi^{q}$, $V(\phi) = V_{0}\phi^{p}$ and $F(\phi, \mathcal{G}) = f_{0}\phi^{w}\mathcal{G}^{k}$, with 
\begin{equation}
    q = \dfrac{w-4k+p(2k-1)}{2k}.
\end{equation}
The Lagrangian \eqref{initiallagr} thus becomes:
\begin{equation}
\mathcal{L} = a^{3}[\omega_{0}\phi^{q} \dot{\phi}^{2}-f_{0}(k-1)\mathcal{G}^k \phi^{w} ]-8f_{0}k \dot{a}^{3} \mathcal{G}^{k-2} \phi^{w-1} [(k-1) \phi \dot{\mathcal{G}}+w \mathcal{G} \dot{\phi}] + a^{3}V_{0}\phi^{p}
\end{equation}
and it satisfies the  existence condition of Noether symmetry  provided that the infinitesimal generators take the form
\begin{equation}
\begin{split}
    &\alpha(a) = -\dfrac{1}{3}[\xi_{0}+kp\gamma_{0}]a, \hspace{0.3cm} \beta(\mathcal{G}) = -4\xi_{0}\mathcal{G}, \hspace{0.3cm} \gamma(\phi) = \gamma_{0}\phi, \hspace{0.3cm}\xi(t) = - \dfrac{(p-w)\gamma_{0}}{4k}t.
\end{split}
\label{eq:sol3}
\end{equation}
A possible solution to the field equations consists of exponential scale factor and scalar field of the form:
\begin{equation}
    a(t) = a_{0}e^{mt}, \hspace{0.3 cm} \phi(t) = \phi_{0}e^{nt}, \hspace{0.3 cm} \G = 24 m^4,
\end{equation}
with additional constraints given by the following system of algebraic equations:
\begin{equation}
\label{algebraicsystem}
    \begin{cases}
    f_{0}(24m^4)^k [k n^2 w^2 + 2 k m n w - 3 m^2(k-1)] +3 m^2 n^2 \omega_{0}+3m^{2}V_{0} = 0\\
    w = q+2\\
    p=w\\
    f_{0} (24m^{4})^{k} w-n \omega_{0} (6 m+n (q+2)) +p V_{0} =0\\
    f_{0} (24m^{4})^k [knw-m (k-1)]- mn^2 \omega_{0}+ mV_{0} =0\,.\\
    \end{cases}
\end{equation}
By replacing the second and the third equation into Eq. (\ref{eq:sol3}), the latter becomes:
\begin{equation}
    \begin{split}
    &\alpha(a) = -\dfrac{kw\gamma_{0}}{3}a, \hspace{0.3cm} \beta(\mathcal{G}) = 0, \hspace{0.3cm} \gamma(\phi) = \gamma_{0}\phi, \hspace{0.3cm}\xi(t) = 0.
\end{split}
\label{generatorsreduced}
\end{equation}
According to the discussion in Sec. \ref{briefintro}, the generators of Eq. \eqref{generatorsreduced} describe an internal symmetry, due to the condition $\xi(t) = 0$. Therefore, symmetries corresponding to this third case can be also selected by the vanishing Lie derivative condition. 
Unlike the other cases, here the system \eqref{algebraicsystem} yields different solutions, but only one of these is completely real. The other solutions, in fact, contain at least one imaginary parameter and have been excluded for obvious physical reasons. For $k=1$ and $k=1/2$, the solution takes the form:

\begin{equation}
k=1/2, \hspace{1cm} n = -\dfrac{3w}{2\omega_{0}}m + \dfrac{\sqrt{m}}{2\omega_{0}}\dfrac{3w^{2}-4w\omega_{0}}{\sqrt{(w^{4}+4\omega_{0}^{2})^{2}-4w^{2}\omega_{0}}}, \hspace{1cm} f_{0} = -\dfrac{\sqrt{6}}{2},
\end{equation}
\begin{equation}
    k=1, \hspace{1cm} n = -\dfrac{6\sqrt{6}w}{\omega_{0}}m^{3} - \dfrac{4 \sqrt{6}w}{\omega_{0}}m^{3}\biggl( \dfrac{3\sqrt{6}m^{2}w^{2}-\omega_{0}}{2\sqrt{6}m^{2}w^{2}-\omega_{0}}\biggr), \hspace{1cm} f_{0} = -\dfrac{\sqrt{6}}{2}.
\end{equation}

\subsection{Fourth case}
\label{fourthcase}
Let us finally analyse the last symmetry selected by the approach, consisting of an exponential kinetic term (\emph{i.e.} $\omega(\phi)=\omega_{0}e^{q\phi}$), a vanishing potential and a coupling function of the form $F(\phi, \mathcal{G}) = f_{0}e^{w\phi}\mathcal{G}^{\frac{1}{2}}$. The corresponding Lagrangian is
\begin{equation}
     \mathcal{L} = 2 \mathcal{G}^{-\frac{3}{2}} f_{0} \dot{a}^3 e^{w \phi} \left(\dot{\mathcal{G}}-2 w \mathcal{G} \dot{\phi}\right)+a^3 \left(\dfrac{1}{2}f_{0} \mathcal{G}^{\frac{1}{2}} e^{w \phi}+\omega_0 e^{q \phi} \dot{\phi}^2\right).
\end{equation}
It is interesting to notice that here the power of $\G$ selected by the Noether Symmetry Approach is not a general constant as the previous cases, but the only coupling function containing symmetries set the value of $k$ to $1/2$. This means that the equivalence with GR in cosmological backgrounds (since $R^2 \sim \G$ in FRW cosmology) naturally arises from symmetry considerations and has not  to be imposed as a requirement. The Noether symmetry existence condition also selects the following infinitesimal generators:

\begin{equation}
\begin{split}
    &\alpha(a) = -\dfrac{\beta_{0}}{12}a, \hspace{0.3cm} \beta(\mathcal{G}) = \beta_{0}\mathcal{G}, \hspace{0.3cm} \gamma(\phi) = 0, \hspace{0.3cm} \xi(t) = -\dfrac{\beta_{0}}{4}t,\\
\end{split}
\label{eq:sol4}
\end{equation}
where, also here, $\xi$ is a function of time. A possible solutions to the equations of motion is given by:
\begin{equation}
    a(t) = a_{0}e^{mt}, \hspace{0.3 cm} \phi(t) = \phi_{0}t,
\end{equation}
with $m, a_0, \phi_0$ being integration constants. In addition, from the Euler-Lagrange equations,  we  get three  constraints on the free parameters, resulting in the following system: 
\begin{equation}
    \begin{cases}
        2 f_{0} m^2 (3 m^2+2 m w \phi_{0}+w^2 \phi_{0}^2)+\sqrt{6} m^{2} \omega_{0} \phi_{0}^2 = 0\\
        w = q\\
        2 \sqrt{6} f_{0} m^{2} w-\omega_{0} \phi_{0} (6 m+w \phi_{0})=0\\
        \sqrt{6} m^{2} \omega_{0} \phi_{0}^2-6 f_{0} m^3 (m+w \phi_{0}) = 0.
    \end{cases}
\end{equation}
To solve the above system, we find $f_{0}$ from the third equation and then we solve the energy condition with respect to $\phi_{0}$, so that we get two different solutions, namely:
\begin{eqnarray}
   && \phi_{0} = -\dfrac{3m}{w}, \hspace{1cm} f_{0} = -\biggl(\dfrac{3}{2}\biggr)^{\frac{3}{2}}\dfrac{\omega_{0}}{w^{2}},
\\
  &&  \phi_{0} = -\dfrac{2m}{w}, \hspace{1cm} f_{0} = -2\sqrt{\dfrac{2}{3}}\dfrac{\omega_{0}}{w^{2}}.   
\end{eqnarray}
The above solutions can be further simplified by choosing  $f_0 = \sqrt{6}/2$. In this way we get:
\begin{eqnarray}
    &&  \phi_{0} = -\dfrac{3m}{w}, \hspace{1cm} f_{0} = \frac{\sqrt{6}}{2}, \hspace{1cm} \omega_0 = \frac{2}{3} w^2,
    \\
    && \phi_{0} = -\dfrac{2m}{w}, \hspace{1cm} f_{0} = \frac{\sqrt{6}}{2}, \hspace{1cm} \omega_0 = \frac{4}{3} w^2.
\end{eqnarray}

\section{Conclusion} \label{sec:conclusions}
We outlined the main properties of the Noether Symmetry Approach discussing some applications and showing how to use the Noether theorem as a method to select theories containing symmetries. Starting from canonical Lagrangians, we first introduced the prescriptions  aimed at constraining the generator of the symmetry. This approach is particularly useful in modified theories of gravity \cite{Bajardi:2022ypn}, or in quantum cosmology, where the conserved quantity allows to restrict the variables superspace to integrable  minisuperspaces \cite{Lambiase, Capozziello:2022vyd}. The first prolongation of the Noether vector yields a more general class of symmetries, which reduce to those coming from the application of vanishing Lie derivative when the functions determining prolongations are  $g=const$ and $\xi=0$. Moreover, an important result arises from the application of Noether symmetry approach to canonical Lagrangians: the infinitesimal generator $\xi$ related to spacetime translations turns out to be a function of the sole time. This result permits to further simplify the approach by reducing the system of differential equations. 

This feature becomes evident when applying the prescription to modified theories of gravity; specifically, in the second part of the manuscript, we focused on a modified gravitational action containing both the Gauss-Bonnet topological term and a dynamical scalar field. The selection of viable models has been pursued by searching for symmetries. In other words, viable modified theories of gravity can be selected adopting a physical criterium based on symmetry considerations. 

Each symmetry  can be related to a conserved quantity that can thus be used to reduce  dynamical equations. As  output of the process, viable cosmological solutions  are  derived  as power-law or de Sitter behaviors. In this perspective, the Noether Symmetry Approach results a method capable of selecting  physically viable models.

\section*{Acknowledgements} \label{sec:acknowledgements}

The Authors acknowledge the support of {\it Istituto Nazionale di Fisica Nucleare} (INFN) ({\it iniziative specifiche}   QGSKY and GINGER). 
This paper is based upon work from the COST Action CA21136, \textit{Addressing observational tensions in cosmology with systematics and fundamental physics} (CosmoVerse) supported by COST (European Cooperation in Science and Technology).

%\appendix*
%\input{sections/appendix1.tex}


\begin{thebibliography}{99}
%\cite{Misner:1973prb}
\bibitem{Misner:1973prb}
C.~W.~Misner, K.~S.~Thorne and J.~A.~Wheeler,
%``Gravitation,''
W. H. Freeman, 1973,
ISBN 978-0-7167-0344-0, 978-0-691-17779-3


%\cite{LIGOScientific:2016aoc}
\bibitem{LIGOScientific:2016aoc}
B.~P.~Abbott \textit{et al.} [LIGO Scientific and Virgo],
%``Observation of Gravitational Waves from a Binary Black Hole Merger,''
Phys. Rev. Lett. \textbf{116} (2016) no.6, 061102

%\cite{EventHorizonTelescope:2019dse}
\bibitem{EventHorizonTelescope:2019dse}
K.~Akiyama \textit{et al.} [Event Horizon Telescope],
%``First M87 Event Horizon Telescope Results. I. The Shadow of the Supermassive Black Hole,''
Astrophys. J. Lett. \textbf{875} (2019), L1

%\cite{EventHorizonTelescope:2019pgp}
\bibitem{EventHorizonTelescope:2019pgp}
K.~Akiyama \textit{et al.} [Event Horizon Telescope],
%``First M87 Event Horizon Telescope Results. V. Physical Origin of the Asymmetric Ring,''
Astrophys. J. Lett. \textbf{875} (2019) no.1, L5

%\cite{Will:2014kxa}
\bibitem{Will:2014kxa}
C.~M.~Will,
%``The Confrontation between General Relativity and Experiment,''
Living Rev. Rel. \textbf{17} (2014), 4

%\cite{Nojiri:2017ncd}
\bibitem{Nojiri:2017ncd}
S.~Nojiri, S.~D.~Odintsov and V.~K.~Oikonomou,
%``Modified Gravity Theories on a Nutshell: Inflation, Bounce and Late-time Evolution,''
Phys. Rept. \textbf{692} (2017), 1-104

%\cite{Odintsov:2023weg}
\bibitem{Odintsov:2023weg}
S.~D.~Odintsov, V.~K.~Oikonomou, I.~Giannakoudi, F.~P.~Fronimos and E.~C.~Lymperiadou,
%``Recent Advances on Inflation,''
[arXiv:2307.16308 [gr-qc]].

%\cite{ParticleDataGroup:2012pjm}
\bibitem{ParticleDataGroup:2012pjm}
J.~Beringer \textit{et al.} [Particle Data Group],
%``Review of Particle Physics (RPP),''
Phys. Rev. D \textbf{86} (2012), 010001

%\cite{Bosma:1981zz}
\bibitem{Bosma:1981zz}
A.~Bosma,
%``21-cm line studies of spiral galaxies. 2. The distribution and kinematics of neutral hydrogen in spiral galaxies of various morphological types.,''
Astron. J. \textbf{86} (1981), 1825

%\cite{Frieman:2008sn}
\bibitem{Frieman:2008sn}
J.~Frieman, M.~Turner and D.~Huterer,
%``Dark Energy and the Accelerating Universe,''
Ann. Rev. Astron. Astrophys. \textbf{46} (2008), 385-432

%\cite{SupernovaSearchTeam:1998fmf}
\bibitem{SupernovaSearchTeam:1998fmf}
A.~G.~Riess \textit{et al.} [Supernova Search Team],
%``Observational evidence from supernovae for an accelerating universe and a cosmological constant,''
Astron. J. \textbf{116} (1998), 1009-1038

%\cite{Vachaspati:2006ki}
\bibitem{Vachaspati:2006ki}
T.~Vachaspati, D.~Stojkovic and L.~M.~Krauss,
%``Observation of incipient black holes and the information loss problem,''
Phys. Rev. D \textbf{76} (2007), 024005

%\cite{Barcelo:2009tpa}
\bibitem{Barcelo:2009tpa}
C.~Barcel\'o, S.~Liberati, S.~Sonego and M.~Visser,
%``Black Stars, Not Holes,''
Sci. Am. \textbf{301} (2009) no.4, 38-45

%\cite{Goroff:1985th}
\bibitem{Goroff:1985th}
M.~H.~Goroff and A.~Sagnotti,
%``The Ultraviolet Behavior of Einstein Gravity,''
Nucl. Phys. B \textbf{266} (1986), 709-736

%\cite{Birrell:1982ix}
\bibitem{Birrell:1982ix}
N.~D.~Birrell and P.~C.~W.~Davies,
%``Quantum Fields in Curved Space,''
Cambridge Univ. Press, 1984,
ISBN 978-0-521-27858-4, 978-0-521-27858-4

%\cite{Weinberg:1988cp}
\bibitem{Weinberg:1988cp}
S.~Weinberg,
%``The Cosmological Constant Problem,''
Rev. Mod. Phys. \textbf{61} (1989), 1-23

%\cite{Niedermaier:2006wt}
\bibitem{Niedermaier:2006wt}
M.~Niedermaier and M.~Reuter,
%``The Asymptotic Safety Scenario in Quantum Gravity,''
Living Rev. Rel. \textbf{9} (2006), 5-173

%\cite{Percacci:2007sz}
\bibitem{Percacci:2007sz}
R.~Percacci,
%``Asymptotic Safety,''
[arXiv:0709.3851 [hep-th]].

%\cite{Bajardi:2021lwp}
\bibitem{Bajardi:2021lwp}
F.~Bajardi, F.~Bascone and S.~Capozziello,
%``Renormalizability of alternative theories of gravity: differences between power counting and entropy argument,''
Universe \textbf{7} (2021) no.5, 148

%\cite{Alexandrov:2002br}
\bibitem{Alexandrov:2002br}
S.~Alexandrov and E.~R.~Livine,
%``SU(2) loop quantum gravity seen from covariant theory,''
Phys. Rev. D \textbf{67} (2003), 044009

%\cite{Maldacena:1997re}
\bibitem{Maldacena:1997re}
J.~M.~Maldacena,
%``The Large N limit of superconformal field theories and supergravity,''
Adv. Theor. Math. Phys. \textbf{2} (1998), 231-252

%\cite{Nilles:1983ge}
\bibitem{Nilles:1983ge}
H.~P.~Nilles,
%``Supersymmetry, Supergravity and Particle Physics,''
Phys. Rept. \textbf{110} (1984), 1-162

%\cite{Rovelli:1997yv}
\bibitem{Rovelli:1997yv}
C.~Rovelli,
%``Loop quantum gravity,''
Living Rev. Rel. \textbf{1} (1998), 1

%\cite{Horava:2009uw}
\bibitem{Horava:2009uw}
P.~Horava,
%``Quantum Gravity at a Lifshitz Point,''
Phys. Rev. D \textbf{79} (2009), 084008

%\cite{Rubakov:1983bb}
\bibitem{Rubakov:1983bb}
V.~A.~Rubakov and M.~E.~Shaposhnikov,
%``Do We Live Inside a Domain Wall?,''
Phys. Lett. B \textbf{125} (1983), 136-138

%\cite{Bajardi:2021hya}
\bibitem{Bajardi:2021hya}
F.~Bajardi, D.~Vernieri and S.~Capozziello,
%``Exact solutions in higher-dimensional Lovelock and AdS$_{5}$ Chern-Simons gravity,''
JCAP \textbf{11} (2021) no.11, 057

%\cite{Qiang:2009fu}
\bibitem{Qiang:2009fu}
L.~e.~Qiang, Y.~Gong, Y.~Ma and X.~Chen,
%``Cosmological Implications of 5-dimensional Brans-Dicke Theory,''
Phys. Lett. B \textbf{681} (2009), 210-213

%\cite{Rasouli:2022tmc}
\bibitem{Rasouli:2022tmc}
S.~M.~M.~Rasouli, S.~Jalalzadeh and P.~Moniz,
%``Noncompactified Kaluza\textendash{}Klein Gravity,''
Universe \textbf{8} (2022) no.8, 431

%\cite{Bajardi:2022ypn}
\bibitem{Bajardi:2022ypn}
F.~Bajardi and S.~Capozziello,
{\it ``Noether Symmetries in Theories of Gravity,''}
Cambridge University Press, Cambridge 2022,\\
ISBN 978-1-00-920872-7, 978-1-00-920874-1

%\cite{Capozziello:2011et}
\bibitem{Capozziello:2011et}
S.~Capozziello and M.~De Laurentis,
%``Extended Theories of Gravity,''
Phys. Rept. \textbf{509} (2011), 167-321

%\cite{Sotiriou:2008rp}
\bibitem{Sotiriou:2008rp}
T.~P.~Sotiriou and V.~Faraoni,
%``f(R) Theories Of Gravity,''
Rev. Mod. Phys. \textbf{82} (2010), 451-497

%\cite{Mishra:2018tqo}
\bibitem{Mishra:2018tqo}
A.~K.~Mishra, M.~Rahman and S.~Sarkar,
%``Black Hole Topology in $f(R)$ Gravity,''
Class. Quant. Grav. \textbf{35} (2018) no.14, 145011

%\cite{Blazquez-Salcedo:2017txk}
\bibitem{Blazquez-Salcedo:2017txk}
J.~L.~Bl\'azquez-Salcedo, F.~S.~Khoo and J.~Kunz,
%``Quasinormal modes of Einstein-Gauss-Bonnet-dilaton black holes,''
Phys. Rev. D \textbf{96} (2017) no.6, 064008

%\cite{Stelle:1976gc}
\bibitem{Stelle:1976gc}
K.~S.~Stelle,
%``Renormalization of Higher Derivative Quantum Gravity,''
Phys. Rev. D \textbf{16} (1977), 953-969

%\cite{Halliwell:1986ja}
\bibitem{Halliwell:1986ja}
J.~J.~Halliwell,
%``Scalar Fields in Cosmology with an Exponential Potential,''
Phys. Lett. B \textbf{185} (1987), 341

%\cite{Uzan:1999ch}
\bibitem{Uzan:1999ch}
J.~P.~Uzan,
%``Cosmological scaling solutions of nonminimally coupled scalar fields,''
Phys. Rev. D \textbf{59} (1999), 123510

%\cite{Capozziello:2002rd}
\bibitem{Capozziello:2002rd}
S.~Capozziello,
%``Curvature quintessence,''
Int. J. Mod. Phys. D \textbf{11} (2002), 483-492

%\cite{Cvetkovic:2016ios}
\bibitem{Cvetkovic:2016ios}
B.~Cvetkovi\'c and D.~Simi\'c,
%``5D Lovelock gravity: new exact solutions with torsion,''
Phys. Rev. D \textbf{94} (2016) no.8, 084037

%\cite{Zanelli:2005sa}
\bibitem{Zanelli:2005sa}
J.~Zanelli,
%``Lecture notes on Chern-Simons (super-)gravities. Second edition (February 2008),''
[arXiv:hep-th/0502193 [hep-th]].

%\cite{Comelli:2005tn}
\bibitem{Comelli:2005tn}
D.~Comelli,
%``Born-Infeld type gravity,''
Phys. Rev. D \textbf{72} (2005), 064018

%\cite{Bajardi:2022tzn}
\bibitem{Bajardi:2022tzn}
F.~Bajardi and R.~D'Agostino,
%``Late-time constraints on modified Gauss-Bonnet cosmology,''
Gen. Rel. Grav. \textbf{55} (2023) no.3, 49


%\cite{Capozziello:1996bi}
\bibitem{cimento}
S.~Capozziello, R.~De Ritis, C.~Rubano and P.~Scudellaro,
%``Noether symmetries in cosmology,''
Riv. Nuovo Cim. \textbf{19N4} (1996), 1-114
%doi:10.1007/BF02742992
%132 citations counted in INSPIRE as of 29 Jul 2023


%\cite{Bajardi:2020osh}
\bibitem{Bajardi:2020osh}
F.~Bajardi and S.~Capozziello,
%``$f(\mathcal {G})$ Noether cosmology,''
Eur. Phys. J. C \textbf{80} (2020) no.8, 704

%\cite{Dialektopoulos:2018qoe}
\bibitem{Dialektopoulos:2018qoe}
K.~F.~Dialektopoulos and S.~Capozziello,
%``Noether Symmetries as a geometric criterion to select theories of gravity,''
Int. J. Geom. Meth. Mod. Phys. \textbf{15} (2018) no.supp01, 1840007

%\cite{Urban:2020lfk}
\bibitem{Urban:2020lfk}
Z.~Urban, F.~Bajardi and S.~Capozziello,
%``The Noether\textendash{}Bessel-Hagen symmetry approach for dynamical systems,''
Int. J. Geom. Meth. Mod. Phys. \textbf{17} (2020) no.14, 2050215

%\cite{Capozziello:2007wc}
\bibitem{Capozziello:2007wc}
S.~Capozziello, A.~Stabile and A.~Troisi,
%``Spherically symmetric solutions in f(R)-gravity via Noether Symmetry Approach,''
Class. Quant. Grav. \textbf{24} (2007), 2153-2166

%\cite{Capozziello:1999xs}
\bibitem{Capozziello:1999xs}
S.~Capozziello and G.~Lambiase,
%``Higher order corrections to the effective gravitational action from Noether symmetry approach,''
Gen. Rel. Grav. \textbf{32} (2000), 295-311

%\cite{Basilakos:2011rx}
\bibitem{Basilakos:2011rx}
S.~Basilakos, M.~Tsamparlis and A.~Paliathanasis,
%``Using the Noether symmetry approach to probe the nature of dark energy,''
Phys. Rev. D \textbf{83} (2011), 103512

%\cite{Capozziello:1996ay}
\bibitem{Capozziello:1996ay}
S.~Capozziello, G.~Marmo, C.~Rubano and P.~Scudellaro,
%``Noether symmetries in Bianchi universes,''
Int. J. Mod. Phys. D \textbf{6} (1997), 491-503

%\cite{Capozziello:2012hm}
\bibitem{Capozziello:2012hm}
S.~Capozziello, M.~De Laurentis and S.~D.~Odintsov,
%``Hamiltonian dynamics and Noether symmetries in Extended Gravity Cosmology,''
Eur. Phys. J. C \textbf{72} (2012), 2068

%\cite{Bajardi:2023vcc}
\bibitem{Bajardi:2023vcc}
F.~Bajardi and S.~Capozziello,
%``Minisuperspace quantum cosmology in f(Q) gravity,''
Eur. Phys. J. C \textbf{83} (2023) no.6, 531

%\cite{Capozziello:2022vyd}
\bibitem{Capozziello:2022vyd}
S.~Capozziello and F.~Bajardi,
%``Minisuperspace Quantum Cosmology in Metric and Affine Theories of Gravity,''
Universe \textbf{8} (2022) no.3, 177

%\cite{Bahamonde:2016grb}
\bibitem{Bahamonde:2016grb}
S.~Bahamonde and S.~Capozziello,
%``Noether Symmetry Approach in $f(T,B)$ teleparallel cosmology,''
Eur. Phys. J. C \textbf{77} (2017) no.2, 107

%\cite{Bajardi:2019zzs}
\bibitem{Bajardi:2019zzs}
F.~Bajardi, K.~F.~Dialektopoulos and S.~Capozziello,
%``Higher Dimensional Static and Spherically Symmetric Solutions in Extended Gauss\textendash{}Bonnet Gravity,''
Symmetry \textbf{12} (2020) no.3, 372

%\cite{Bahamonde:2019swy}
\bibitem{Bahamonde:2019swy}
S.~Bahamonde, K.~Dialektopoulos and U.~Camci,
%``Exact Spherically Symmetric Solutions in Modified Gauss\textendash{}Bonnet Gravity from Noether Symmetry Approach,''
Symmetry \textbf{12} (2020) no.1, 68

%\cite{Bahamonde:2019jkf}
\bibitem{Bahamonde:2019jkf}
S.~Bahamonde and U.~Camci,
%``Exact Spherically Symmetric Solutions in Modified Teleparallel gravity,''
Symmetry \textbf{11} (2019) no.12, 1462

%\cite{Agrawal:2020xek}
\bibitem{Agrawal:2020xek}
P.~Agrawal, S.~Gukov, G.~Obied and C.~Vafa,
%``Topological Gravity as the Early Phase of Our Universe,''
[arXiv:2009.10077 [hep-th]].

\bibitem{Lambiase}
S.~Capozziello and G.~Lambiase,
%``Selection rules in minisuperspace quantum cosmology,''
Gen. Rel. Grav. \textbf{32} (2000), 673-696

\end{thebibliography}
\end{document}